\documentclass[12pt]{article}

\renewcommand{\bar}{\overline}

\begin{document}
\begin{flushright}
SLAC--PUB--9000\\
November 2001
\end{flushright}
\bigskip\bigskip
\bigskip\bigskip
\baselineskip 21pt

\centerline{{\bf Summary and Outlook for 9th International Symposium on Heavy Flavor Physics}
    \footnote{\baselineskip=14pt
     Work supported by the Department of Energy, contract
     DE--AC03--76SF00515.}}
\vspace{22pt}
  \centerline{\bf Helen Quinn}
\vspace{8pt}
  \centerline{\it Stanford Linear Accelerator Center}
  \centerline{\it Stanford University, Menlo Park, California 94025}
\vspace*{2cm}
\begin{center}
Abstract
\end{center}

This is the summary talk of a meeting held at the California
Institute of Technology Sept 10--13, 2001. I do not attempt to
summarize all the beautiful experimental results we have seen this
week, nor to repeat the lively theoretical discussions that have
occurred. Rather I will present my own biased perspective on what
we have learned, and on the important tasks that need our
attention as we work to make the most of the rapidly accumulating
data in this field. \vspace*{2cm}

  \centerline{Talk presented at 9th International Symposium }
  \centerline{on Heavy Flavor Physics}
  \centerline{California Institute of Technology}
  \centerline{Pasadena, CA}
  \centerline{September 10-13, 2001}
\vspace*{0.9cm}
\newpage

\setlength{\baselineskip}{15pt}

\medskip\leftline{\bf INTRODUCTION}

I want to begin by thanking our hosts for presenting a
well-organized and very effective meeting in a week where such
tasks were not easy. Given the dreadful events of last Tuesday we
were all a little distracted. None of us will forget where we were
in this week, nor that we were well taken care of here. Despite
the disaster (which, as one small side-effect, shut down CalTech
for the day) a new location for our meeting on Tuesday was found,
new arrangements to feed us made quickly and smoothly, and our
meeting carried on. The technology continued to function, the
talks were all given. We all appreciate the excellent hospitality
we were given throughout the week, and the support we are now
offered as we all try to figure out how we will get home.  We owe
a vote of thanks to David Hitlin, Frank Porter, Gregory
Dubois-Felsmann, and Anders Ryd and their support staff for their
efforts. They made it look easy, but I am sure it was not.

To turn to the physics, I start with a personal comment. I have
greatly enjoyed being part of the process of development of the
$B$ factory and the BaBar experiment almost from the beginning.
This was a fascinating and humbling experience for a theorist. I
have seen how much the naive first estimates of what such a
facility can achieve are transformed as the reality of building an
actual experiment and analyzing real data take over from the
back-of-the-envelope guesses with which we begin the process. On
both the theory side and the experimental side we (and from here
on we means not just BaBar collaborators but all working in this
field) have learned where our first approximations are
insufficient. The process could be disheartening, except that
there are enough clever and determined people involved that
somehow we continue to make progress. Despite the difficulties, we
manage to maintain optimism and work hard enough that real
progress is made. The results on both the theory side and the
experimental side that we have seen this week are a proof of that.
The job is far from done, but it is well begun. We have begun to
unravel the physics of $B$ meson decays.

I mention BaBar first only because it is there that I sit and so I
have seen the process there first hand.  The same hard work and
persistence in other laboratories has also produced new and
interesting results reported this summer. We have heard many
results this week (even a few actually new ones) and over this
last summer. In $B$ physics CLEO, Belle and BaBar are reporting
results on some branching ratios as small as $10^{-6}$
\cite{websites}, and both BaBar and Belle have reported evidence
that the $CP$ violating parameter sin ($2\beta$) (or sin
($2\phi_1$) which is the same thing) is non-zero \cite{sin2beta}.
With current accuracy, the result is consistent with the range
predicted by the Standard Model. Results on charm, strange and tau
physics have also been reported here. I do not intend to summarize
all these results, they speak for themselves in the many excellent
presentations we have heard \cite{proceedings}. Ongoing
experiments will bring more new results and we can expect improved
precision on interesting measurements for some time to come.

So what have we learned from all these results? What more do we
expect to learn by pursuing the physics of flavor over the next
few years? We certainly have learned that persistence pays, and
indeed is required to reach our goals. Both in theory and in
experiment we must continue to work hard, for several more years
at least, to truly pin down all the details of this sector. We
have seen that in $B$-physics as in previous flavor physics there
is no free lunch. The golden channel of $\psi K_s$ and its cousins
($b\rightarrow c \bar c s$ decays of $B_d$) is interesting. We
were not lucky enough to find any challenge to the Standard Model
in this first observation of $CP$ violation in the neutral $B$
meson decays. Since one of our major motivations for pursuing this
physics is to test the Standard Model story on $CP$ violation that
simply means that we must continue with the rest of the program,
implementing physics analyses of many more channels.

In our struggle to understand the physics of flavor and with it
the physics of $CP$ violation, for at least within the Standard
Model the two are intimately linked, we have, once again, run into
the fact that we cannot isolate quarks. This means that to study
their decays we must inevitably also study some aspects of strong
interaction hadronic physics. There are few ``golden channels'' in
which the physics of interest to us for testing the Standard Model
 can be cleanly separated from the hadronic physics.  Our calculational
 tools are then in
need of help. The theory work advances steadily, but the net
result is that the predictions we need to be able to make depend
on quantities that we cannot directly calculate. We must therefore
take a multichannel approach, where theory and experiment feed
information to one another.

Theoretical predictions, once hadronic physics enters the picture,
depend on some inputs that can only be obtained from measurement
or models. Interpretation of measurements depend on input from
theory. The process of learning is murky and iterative. In the
rest of this talk I will illustrate this with some examples from
the theory talks we have heard this week. I will  draw a few
morals about the way this game must be played if it is to succeed.
Being an equal opportunity moralist I have advice for theorists
and for experimentalists alike. In the end it comes down to the
same advice---a plea for logical clarity and honest revelations
about what is an input and what is an output in presenting any
theory work or any experimental result. I will make myself clearer
on this after the examples. I am going to say a lot of obvious
things in this talk, at the risk of sounding preachy and naive.
The  reason I do this is because it seems sometimes that our tools
become so sophisticated that we forget some obvious things. I
guess I'm old enough to get away with being preachy on occasion.

My first example is the measurement of charmless semileptonic $B$
decays and the extraction of $V_{ub}$. Here we heard of new
theoretical work from Christain Bauer \cite{bauer} and ideas on
using the spectrum of $B\rightarrow s\gamma$ data as a way to
measure some input parameters from Ira Rothstein \cite{Rothstein}.
The CLEO experiment has begun to implement some of these ideas
\cite{cleomoments}.  You all know the problem. To extract $V_{ub}$
we must measure charmless decays. The cuts on the data that must
be introduced to remove the background from the much more common
decays to charm bring with them a price for the theorists. The
fraction of the data kept after any cuts must be calculated. This
fraction is sensitive to details of the spectrum that may not be
reliably predicted by a  quark level calculation.

Christian Bauer showed us how a combination of cuts can be used to
limit the sensitivity to this problem and suggested how one can
also gain some tests of that sensitivity by varying the cut
prescription.  He advised minimizing sensitivity to theory. That
is impossible; we are trying to extract a theory parameter. But
what he really meant was minimizing sensitivity to the
uncertainties in the theory that arise from soft physics. That can
be done! I would further suggest that the data should be presented
in two ways. The cuts may be tuned based on theory input, but the
result should first be stated in as theory independent a fashion
as possible. Only after that should the analysis that gives the
theoretical parameter be introduced. In this case this separation
is readily made.
 One has simply to quote a rate of events in a given kinematic region, or
perhaps a table of such numbers for different choices of kinematic
cuts.  The second step, of turning that table into a best estimate
for $V_{ub}$, is also needed. It
 should be kept separate. In this step theory and experiment become
 inextricably mixed together.
The reason I plead for the first step, the presentation of data
with no theory in the numbers, is that that is what will allow us
to come back at a later date (even possibly after the
collaboration presenting the results is long disbanded) and
re-analyze the data with new theory inputs. Experiments should not
become so theory-driven that they only present results for
theoretically interesting quantities. I know there are cases where
my advice is simply too naive, where there is essentially no way
to present a theory-free set of numbers. It is just this fact that
has led us into the bad habit of confounding the two steps,
 that of measurement and that of interpretation. The pattern is perhaps
 reinforced when
experiments publish only as letters; there is not room in a letter
for the two steps, and the second number is regarded as the real
result of the measurement. In the short term this is true, in the
long term the theory independent result has more staying power! My
advice to experiments is to pay attention to whether or not they
are forced into this procrustean bed, or whether they are allowing
themselves to fall there simply because they have not tried to
hard enough avoid it!

I want to return to the idea that Ira Rothstein \cite{Rothstein}
talked about, that of using one set of channels to measure some
soft physics parameters and then using those parameters in
interpreting another set of channels. This may sound obvious, it
is in practice neither obvious nor simple to implement. The issue
is that these parameters are not physical quantities in the
technical sense---they are definition and convention dependent.
This means that one must in fact do some higher order QCD
calculation to understand how the spectrum in say radiative $B$
decays is related to that in semileptonic $B$ decays. This is
ongoing work, and the ``best'' or even standard conventions for
defining the relevant $B$ physics parameters may not yet have
emerged. This is a technical point, not a problem. However when
experiments quote numbers for a theoretically-defined quantity,
such as the parameter $\bar \Lambda$ of the heavy quark effective
theory, they must at the same time tell us what definition of this
parameter (or equivalently, in this case, of quark mass) they are
using. Without such a definition the quantity is meaningless.

A second topic where we saw some interesting theoretical analysis
was on the subject of $D \bar D$ mixing, long advertized by
theorists as a good place to look for new physics effects. In the
talk by Zoltan Ligeti \cite{ligeti} we heard how the naively
predicted pattern of the operator product expansion contributions
to $x$ and $y$ is possibly disrupted because of SU(3) breaking
effects. Since the otherwise-leading operator is suppressed by
SU(3) symmetry, the symmetry-breaking effects can be significant.
Indeed they may alter the expected relative sizes of $x$ and $y$.
The size of expected Standard Model effects is also enhanced once
SU(3) breaking is considered. So the lesson here is not to trust
first rough estimates of theoretical uncertainties, these things
always some study. However the effects here are still at most at
the few percent level, so this remains a place to look for new
physics effects.

No method can completely remove the issues of theoretical
uncertainties. No matter what technical improvements are made, in
the end we must are rely on some quark-level calculations and some
version of quark-hadron duality. Now it is the job of the
theorists to estimate how big the remaining uncertainties could
be. Unfortunately there is often no rigorous approach that gives a
definite answer. I believe that theorists must give skeptical
answers here, but that it is important to try to be quantitative.
It is just as bad to give an off-the-cuff overestimate of the
uncertainties as it is to claim that everything is under control
with an optimistic estimate of uncertainties.

So let us spend a little time considering how big can the
violations of quark hadron duality be?  What sets the scale of
these effects? What level of averaging is needed to remove the
detailed dependence on resonance masses in a spectrum and hence to
get results that depend only on the underlying quark diagrams? To
make this point I will digress a little and talk about a case
where I know something about the answers, that is the rate of
hadron production in $e^+e^-$ annihilation. This is typically
expressed as a ratio $R_{e^+e^-}$ of hadron to non-resonant
muon-pair production \cite{smearing}.  One can prove quite
formally that an integral over this ratio along the entire real
physical cut can be given by the relevant quark-loop graph
calculated in the deep Euclidean region. Indeed one can use the
analytic properties of the quark graph to calculate an integral
that weights a small segment of the cut. In the Euclidean region
there are no small denominators from near-mass shell propagators
and so the power series expansion is well behaved.

What limits the reliability of the calculation is that as one
approaches a threshold on the physical cut certain propagators in
the diagram give small denominator factors. We know which diagrams
have to be summed to all orders to produce the onium resonances
just below a new quark threshold. Right on resonance the
violations of duality are huge if we do not make this resummation.
We can avoid the resummation only if we calculate an integral that
can be determined without the knowledge of the function close to
the threshold.  That is the essence of the ``global'' duality --it
gives us a correct averaged cross-section, but not the detailed
threshold and sub-threshold structure. On the other hand, once we
are well above threshold we can approach quite close to the
physical cut without any small denominators appearing in any
diagram. Indeed we find that the averaged rate and the naive
``local" quark calculation agree well in such a region. There are
two main points that this example teaches us.  The first is that
the violations of local duality are very large if we are foolish
enough to try to use it where some quark propagator goes on-shell,
but quite small when we are far from any threshold.  In a properly
averaged quantity, the ``duality violations'' are small. The
second point is that a careful enough examination of the quark
diagrams revealed what went wrong with the perturbation series at
the threshold, and hence indicated the range over which averaging
was needed to avoid these problems. The size of duality violations
is then controlled by the averaging scale compared to
$\Lambda_{QCD}$ (and also to light quark masses which are
smaller).

These features then generalize to two questions. The first is how
much do we need to average over hadronic mass spectra to ensure
that corrections to a quark-hadron duality estimate are under
control? The second is to find what effects can give large
violations for the process in question, and what sets the scale of
these effects in a suitably averaged quantity. In an ideal world
we could give rigorous answers to these questions, in the real
world different theorists reach different conclusions. For example
for the case of the inclusive semileptonic decay rate to particle
with charm, where no significant kinematic cuts are required to
select events, these corrections have been discussed in some
detail and are generally agreed to be be small. The range of
lepton momenta achieves some averaging over the hadron mass
spectrum.  The argument among theorists is then over whether small
means a few percent or of order $10^{-3}$. Bigi and Uraltsev have
argued that the latter number is appropriate \cite{vademecum}.
Their argument is plausible; no one has shown specific effects
which they have ignored. However this same type of argumentation
leads to other results that are discrepant with experiment, for
example the differences in the lifetimes of different
$b$-containing mesons and baryons are larger than duality-based
arguments would predict. The conundrum is then whether we are at
the stage where we should say these are serious violations of
Standard Model predictions, or simply that violations of
quark-hadron duality are bigger than expected. Any skeptical
observer takes the latter position. But the results certainly show
how important it is to try to gain a better understanding of how
to quantify such duality violations reliably. We may lose the
ability to recognize many signals of new physics if we do not do
better on this front. Theorists generally divide into those who
have made the calculations and are willing to make estimates of
the size of the breaking, and skeptics who say we cannot reliably
estimate these effects. The skeptics typically cannot point to any
specific error in the calculations that have been made. I admit
that I have generally been among the latter group---but at this
stage of the game it is an inadequate position to take. We
theorists need to keep attempting to do better.

However, at the same time as I preach that skeptical theorists
should take on this challenge, I warn the experimentalists that
the most aggressive (optimistic) estimates of theoretical
uncertainties are often not reliable. The problem is that the
diseases of the quark-level perturbation theory calculation can be
very subtle and do not always manifest themselves in an obvious
way.  For the case of $B$ decays, we will only learn how to limit
the size of duality violating effects by comparing predictions
with data in many, many channels. In particular it is important to
test the sensitivity of results to variation of any experimental
cuts or changes in procedure. When there is significant
sensitivity in the parameter extraction to cut-variation or to
input models or assumptions then we know that we cannot trust the
duality-based calculation. However, even when the sensitivity to
variation in input is small we still cannot be certain that the
effects of duality violations are small. This all leaves a very
murky path to finding a distinction between new physics effects
and unexpectedly large corrections to our calculations. I hope and
expect that we will be able to find ways to separate these two
things, but there are no guarantees.

One class of processes for which this process is now at a
well-developed stage is the calculation of two-pseudo-scalar decay
channels using qcd-improved factorization or the alternative
formulation called perturbative qcd. We have all seen the papers
and heard talks by two groups working on this topic. It has taken
me some time to understand some of the details of this work.  The
results of the two groups are quite different, but it appears at
first glance that they are pursuing the same methods. The methods
both keep leading order terms in $\Lambda_{QCD}/m_b$ and calculate
the leading $\alpha_s(m_b)$ corrections. The groups get quite
different results on some points. The problem is that even the
conclusions on the power counting for the relative sizes of
certain contributions are dependent on assumptions about the quark
distribution function end-point behavior. In my opinion the fact
that two independent groups have been working on this is
important. We have seen that the estimates of theoretical
uncertainties of one group have evolved due to the work of the
other group. This is what we need, honest efforts to give good
error estimates, and, to find what parts of those estimates are
sensitive to assumptions, more than one group doing independent
work on the same problem. This then leads to an iterative process
that can eventually give us some confidence that we have a good
estimate of theoretical uncertainties.

The group of Beneke {\em et al.} \cite{bbns}, represented in this
meeting in the talk by Matthias Neubert, assume that the end point
behavior is a power of $x$, and do not include any Sudakov
suppression factor. They argue this effect does not play a
significant role at the actual $B$ mass scale. They then use data
and models to give the needed input for matrix elements; for
example, the transition matrix element for $B$ to pseudoscalar
meson is taken from semileptonic decay measurements. They make an
honest effort to test a range of assumptions,  for example for
quark distribution functions, and see how their results vary as
those input assumption are varied. This is the model we all must
follow. But no matter how honest a group might be, it is important
to also have another group or groups think independently about
what input assumptions are reasonable. Only in this way can there
be a serious discussion about whether the estimates of theory
uncertainties are conservative or aggressive.

The group of Keum {\em et al.} \cite{qcdpert}, represented here by
the talk by Y-Y. Keum, make a different set of assumptions. They
include transverse-momentum dependence via  Sudakov factor in the
quark distribution functions. This suppresses the contribution
which is the leading term in the Beneke {\em et al.} approach, so
that in this so-called perturbative QCD approach the dominant term
is an order $\alpha_s(m_b)$ hard-gluon exchange to the spectator
quark. Then the inputs are light-cone quark distribution functions
for the mesons, which they argue are calculable. I find that last
statement quite doubtful (see the arguments on this subject in a
paper that appeared shortly after this talk \cite{sachrajda}).
However one can simply regard these inputs as an alternate (and
stronger) set of assumptions. One then must ask whether these are
reasonable assumptions. They choose the parameters of the Sudakov
term so that the scale of average transverse momenta in the quark
distributions is $k_{\perp}^2 = m_b \Lambda_{QCD}$. This is an
incorrect choice! Even for the $B$ meson, and certainly for the
pion, the transverse momenta should be scaled only by
$\Lambda_{QCD}$. This choice has significant implications for
their numerical results, which I therefore find suspect.

However even if their results are not yet reliable, there are some
points that have been raised by these authors that have led to
some revisions of the error estimates of the former group. This
reinforces my statement that it is always useful to have two
groups approaching the problem independently, to test assumptions
about theoretical uncertainties. The major issue raised by the
second set of calculations is the impact of, and control over, the
($\Lambda_{QCD}/m_b$)-suppressed contributions. The contribution
of annihilation graphs is one such term. In the Beneke {\em et
al.} formalism it turns out that this contribution is infra-red
singular and hence dependent on the cut-offs introduced to control
this singularity. Recent papers of this group include an larger
estimate of the  uncertainties due to these terms than earlier
papers (as far as I can tell). In the Keum {\em et al.} approach
the infrared behavior of this contribution is softened by the
$k_{\perp}$ dependence of quark propagators, and the term is found
to be numerically significant. In particular it contributes to a
large imaginary part, the size of which is important for any
estimate of direct $CP$-violation. These numerical results depend
on the incorrect scale for the average $k_{\perp}$ mentioned
above, and so I do not trust them. However the fact that this
contribution needs to be estimated carefully is an important point
that was raised by this work. The public debate between the two
groups has also helped those not directly involved to learn what
are the critical issues.  That is an important impact of having
two sets of calculations.

This example provides a model for our path into the future. It is
not possible to avoid all hadronic physics questions. We need
theorists to tackle them and to make some assumptions in order to
do so. We then must do a serious job of thinking about the range
of input assumptions. Human nature being what it is, we must have
more than one group tackle that job. That helps  ensure a public
discussion of all relevant issues. However  much this is done,
there will always be some remaining tension between those who have
made the estimates and skeptics who doubt their methods. The
history of the $\Delta I= 1/2$ enhancement and ${ \epsilon^\prime
\over \epsilon}$ provide a warning that favors the skeptics, but
$B$-decays offer a new regime where more systematic expansions are
available. There is ongoing progress on the theory of this regime.

So this effort must continue, both on the theory front in
estimating uncertainties, and on the experimental front in
presenting measurements in as theory-independent a fashion as
possible. Then the clever application of theory to experimental
data is needed, to provide the cleanest possible tests of the
Standard Model. That requires close cooperation of theorists and
experimentalists. Even after all the hard work is done the
question remains whether we can be confident enough of our
uncertainty estimates that we can be sure a discrepant measurement
tells us there is new physics. The pessimistic view is that we
will always be able to adjust the theory uncertainties to cover
any measured results. The optimistic view is that eventually we
will gain enough confidence in our methods to recognize  true
discrepancies if such exist. I suspect that any one result will
not be convincing. It will take a pattern of discrepancies in
fitting many channels to tell us that the Standard Model is
failing. Both theorists and experiments still have much work to
do. I have been impressed up till now  by the persistence on both
fronts in this work. This persistence has yielded steady progress,
as we have seen in this meeting. I am hopeful that this progress
will continue.

\end{document}